\documentclass[12pt]{article}
\usepackage{epsfig,amssymb}

\newcommand{\be}{\begin{equation}}
\newcommand{\ee}{\end{equation}}
\newcommand{\ba}{\begin{eqnarray}}
\newcommand{\ea}{\end{eqnarray}}
\newcommand{\bq}{\begin{equation}}
\newcommand{\eq}{\end{equation}}
\newcommand{\bqa}{\begin{eqnarray}}
\newcommand{\eqa}{\end{eqnarray}}
\newcommand{\ben}{\begin{enumerate}}
\newcommand{\een}{\end{enumerate}}
\newcommand{\bc}{\begin{center}}
\newcommand{\ec}{\end{center}}
\newcommand{\bqb}{\begin{eqnarray*}}
\newcommand{\eqb}{\end{eqnarray*}}

\def\pr#1#2#3{ Phys. Rev. ${\bf{#1}}$, #2 (#3)}

\def\pl#1#2#3{ Phys. Lett. ${\bf{#1}}$, #2 (#3)}
\def\prep#1#2#3{ Phys. Rep. ${\bf{#1}}$, #2 (#3)}

\def\np#1#2#3{ Nucl. Phys. ${\bf{#1}}$, #2 (#3)}

\def\etal{{\it et.al.\/}}

\global\nulldelimiterspace = 0pt

\begin{document}
\pagenumbering{arabic}
\thispagestyle{empty}
\def\thefootnote{\fnsymbol{footnote}}
\setcounter{footnote}{1}

\begin{flushright}
PM/99-28 \\
July 1999\\
corrected version \\

 \end{flushright}
\vspace{2cm}
\begin{center}
{\Large\bf 
The role of the top mass in $b$-production}\\ 
{\Large\bf
at future lepton
colliders}\\
\vspace*{0.5cm}
{\large 
M. Beccaria$^{(1)}$, 
P. Ciafaloni$^{(2)}$, 
D. Comelli$^{(3)}$, \\
F. Renard$^{(4)}$,
C. Verzegnassi$^{(1)}$}\\
\vspace{0.7cm}
${}^1$ 
Dipartimento di Fisica dell'Universit\`a di Lecce, I-73100, Italy,\\
and Istituto Nazionale di Fisica Nucleare, Sezione di Lecce;\\
${}^2$
Istituto Nazionale di Fisica Nucleare, Sezione di Lecce; \\
${}^3$
Istituto Nazionale di Fisica Nucleare, Sezione di Ferrara; \\
${}^4$
Physique
Math\'{e}matique et Th\'{e}orique, UMR 5825\\
Universit\'{e} Montpellier
II,  F-34095 Montpellier Cedex 5.

\vspace{0.2cm}

\vspace*{1cm}

{\bf Abstract}

\end{center}

We compute the one loop contribution coming from vertex and box diagrams, where
virtual top quarks are exchanged, to the asymptotic energy behaviour of
$b\bar b$ pair production at future lepton colliders. We find that
the effect of the top mass is an extra linear logarithmic term of Sudakov type
that is not present in the case of $(u,d,s,c)$ production. This appears
to be particularly relevant in the case of the $b\bar{b}$ cross section.

\def\thefootnote{\arabic{footnote}}
\setcounter{footnote}{0}
\clearpage

A well known feature of $b\bar b$ production at the $Z$ peak is the
fundamental role of the top quark mass. Its effect in the $Zb\bar b$
\underline{vertex} generates a contribution to the partial $Z$ width
into $b\bar b$ pairs $\Gamma_b$ that contains a quadratic term (and,
also, an almost equally important logarithmic term) in the top mass
$\simeq \alpha m^2_t/M^2_Z$ \cite{mtvert}. Numerically, this turns out
to be of a relative few percent size for $m_t\simeq 175~GeV$, well
beyond the experimental accuracy of the related measurement
\cite{expcombi}. Thus, neglecting the top mass in this special vertex
would induce a catastrophic error in the theoretical prediction
\cite{thpred}.\par
One might wonder whether this fondamental top mass role is retained
when one moves away from the $Z$ peak and considers the generalization
of that process, lepton-antilepton into a fermion-antifermion pair, at
higher energies. Here one must make a preliminary observation. Away
from the $Z$ peak, $Z$ exchange is no longer necessarily dominant. In
fact, the photon exchange acquires a certain relevance (in the case
of muon pair production it actually becomes the leading contribution).
This introduces new diagrams to be considered at the one loop level, where
virtual tops can be exchanged. Moreover, the role of box diagrams
appears to be also increasing with energy \cite{box}, and such diagrams
contain virtual top contributions as well. Therefore, one expects in
full generality to find several effects of the top mass at higher energies in
the considered four-fermion processes.\par
The general procedure for computing 1-loop effects in the massive
$f\bar f$ case exists in the literature \cite {Hollik}.
The detailed study done for the special case of $t\bar t$ production
\cite {Hollik} confirms the idea
that the role of the top mass might also be relevant in the process   
of $b\bar b$ production. To verify this feeling, 
several types of contributions,
besides those that were effective at the $Z$ peak,must be carefully
computed.\par
A partial reduction of these effects is provided by the observation
that a special theoretical framework, based on "$Z$-peak subtraction"
\cite{Zsub}, can be used to study these processes. In this approach,
those one-loop effects that appear at the $Z$ resonance are
automatically reabsorbed into new theoretical input data. In this way,
that component of the $Zb\bar b$ vertex whose dependence on $m_t$ is
constant with energy disappears, i.e. is incorporated into the partial
width $\Gamma_b$. Thus, the "effective" $m_t$ dependence, which is in
our approach, we repeat, the one which \underline{cannot} be eliminated by
LEP1, SLC measurements, will be given by those components of
self-energies and vertices whose dependence on the top mass varies with
energy, and by a set of box diagrams, that by construction cannot be
reabsorbed by $Z$-peak measurements.\par
One might ask under which conditions, a priori, such $m_t$ effects should be 
interesting at future $l^+l^-$ colliders. In principle, one would
expect from the $Z$-peak situation effects of the few percent size;
these might be relevant if the experimental accuracy were (at least)
comparable with the remarkable LEP1,SLC level.\par
A further motivation for considering these $m_t$ exchanges is provided, in our 
opinion, by
the conclusions of recent works \cite{CC, sud} where the effects of
asymptotic leading contributions 
(essentially, vertices and boxes with $W$
exchange) is studied for four-fermion processes in the $TeV$ region. As
shown in those papers, at such energies new electroweak effects of
"Sudakov-type" arise that are of linear and squared logarithmic type,
typically $\simeq \ln{q^2\over M^2_W},~\ln^2{q^2\over M^2_W}$ where $q^2$
is the squared c.m. energy. Such terms become separately rather 
large (typically, at the relative ten percent level) in the 
considered energy region, although a cancellation process is 
at work that reduces their overall effect. The analysis of
ref.\cite{sud} was performed in the Feynman-t'Hooft $\xi=1$ gauge,
ignoring systematically possible massive virtual top exchanges in vertices and
boxes, the only diagrams responsible for Sudakov-type effects. Given the fact 
that small
modifications can in principle 
alter the cancellation mechanism, we feel that such
exchanges might have some relevance for what concerns $b$ production in
this energy regime.\par
The purpose of this paper is precisely that of showing to what extent
this feeling is actually correct. 
With this aim, we shall start from a series of
theoretical formulae whose origin can be immediately found in
ref.\cite{sud}, to which we defer in order to make 
this short article not
too much filled with repetitions and definitions.\par
To begin our analysis, we recall briefly what is the origin of the
"Sudakov-type" effects in the considered four-fermion processes.
Briefly, for massless virtual quark exchanges,
in the Feynman-t'Hooft gauge
they are originated from the set of vertex- and box-type
diagrams shown in Fig.1. At high energies,
the double $\ln$'s are generated by vertices with one $W$, $Z$
exchange and by boxes. Linear $\ln$'s are generated by all diagrams
depicted in Fig.1. 
In general, these $\ln$'s add to those that can be associated to the RG
"running" of the various couplings. The latter ones are originated by
self-energy diagrams and by the universal "pinch" \cite{pinch}
component of the vertex with two $W$'s.\par
The asymptotic expansions, valid in a region where $q^2 >> M^2_Z$, of
four-fermion observables was derived in ref.\cite{sud} treating, we
repeat, all
virtual quark exchanges contributing Fig.1, including the top one, as
if all quarks were massless. In other words, the predictions
for the $b\bar b$ production were identical with those for $d\bar d$,
$s\bar s$ quarks.\par
Taking now properly into account the virtual top exchange 
corresponds to performing an accurate estimate of the vertex and box
diagrams shown in Fig.2 in the case of the final $b\bar b$ production.
As one sees, one must add a new set of diagrams that correspond
to charged would-be Goldstone boson $\Phi^{\pm}$ exhanges, whose
contribution in the $\xi=1$ gauge where we are working is
essential. Although the computational procedure is very similar to that 
illustrated in
ref.\cite{sud}, there is one feature that, we feel, deserves the small
dedicated discussion that follows.\par
Let us begin with a consideration of the first diagram of Fig.2. Then we
must discuss separately the two cases of single $W^+$ and of single
$\Phi^+$ (charged would-be Goldstone boson) exchange. We remind the
reader that our calculations are performed in the t'Hooft $\xi=1$
gauge. Since we are dealing systematically and by construction with
gauge-independent combinations of diagrams, we can set
$M_{\Phi^+}=M_{W^+}$ from now on, without prejudice on the total
observable contributions, as discussed exhaustively in
ref.\cite{sud}.\par
The calculation of the asymptotic limit of the diagram with one $W$
leads to the conclusion that, for this case, the two leading
logarithmic terms in the expression for the final $b\bar b$ pairs are
the \underline{same} as those for $d\bar d$, $s\bar s$ pairs. In other
words, there is no extra "Sudakov top" effect in this case. This
conclusion is to a certain extent not unexpected if one remembers that
a completely similar property was valid at the $Z$ peak, where $W$
exchange was not generating any $\simeq m^2_t$ contribution. In fact,
the same property was also characteristic of the diagram with two $W$
exchange at the $Z$ peak, and one would thus expect to find it again in
the next two $W$ diagram of Fig.2.\par
The diagram with one $\Phi^+$ exchange leads to a "genuinely" new top
effect. In fact, it produces a linear logarithmic term that is
proportional to $\simeq \alpha{m^2_t\over M^2_W}\ln{q^2\over m^2_t}$. 
With the notation of ref.\cite{box, Zsub} we find the leading
contribution to the $\gamma$ or $Z-b\bar b$ vertices:
\bq
\Gamma^{V}_{\mu,b}(\Phi^+)=C^V{|e|\alpha\over32\pi s^2_W}
({m^2_t\over M^2_W})\ln({q^2\over m^2_t}) \gamma_{\mu}(1-\gamma^5)
\eq

\noindent
with $C^{\gamma}=Q_t$ and $C^{Z}={-2s_W\over3c_W}$.

Note that, in practice, one could "bargain" $m_t$ with $M_W$ in the
logarithm, and the difference would be embodied by a constant $\ln 
\frac{m_t}{M_W}$, that
would be asymptotically negligible. In the limit $m_t>>M_W$, 
the single W exchange diagram also gives 
terms of order $ln^2(m_t^2/M^2_W)$. We postpone the discussion of
"constant" ($q^2$-independent) contributions to the end of the
paper.
From now on, though, we shall
retain the notation $\simeq \ln{q^2\over m^2_t}$ to remind the origin of
the logarithmic term.\par
The calculation of the second diagram leads to the conclusion that we
have anticipated. The two $W$ exchange leads to the same asymptotic
expansion for either $b$ or $d,s$ pairs. The (one $W$ + one $\Phi$) 
exchanges give contributions which vanish
asymptotically like $m_t^2/q^2$. 
The two $\Phi$ exchange leads
to a linear logarithmic term, again of the form $\simeq
\alpha{m^2_t\over m^2_W}\ln{q^2\over m^2_t}$. 

\bq
\Gamma^{V}_{\mu,b}(\Phi^+\Phi^-)=C^V{-|e|\alpha\over32\pi s^2_W}
({m^2_t\over M^2_W})\ln({q^2\over m^2_t}) \gamma_{\mu}(1-\gamma^5)
\eq

\noindent
with $C^{\gamma}=1$ and $C^{Z}={1-2s^2_W\over2s_Wc_W}$.

Finally, we consider
the box diagram
with two $W$ exchange. Here there is no $Z$-peak analogue, which
prevents us from having a related theoretical prejudice. In fact, an
accurate calculation leads to the conclusion that, again, there are no
extra logarithmic terms with respect to the $d,s$ cases, exactly like
in the $W$, two $W$ vertex situations.\par
Keeping in mind the previous discussion, and following the same
procedure as in ref.\cite{sud}, we can now write the
"corrected" expressions for the various observables containing a
final $b\bar b$ pair.\par
Using $m_t=175~ GeV$, the theoretical expansion of
$\sigma_b$, the cross section for $b\bar b$ production is given by:
\bqa
\sigma_{b}&=&\sigma^{B}_{b}[1+{\alpha\over4\pi}\{(10.88N-53.82
)\ln{q^2\over\mu^2}\nonumber\\
&&+(76.75 \ln{q^2\over M^2_W}+11.98\ln{q^2\over M^2_Z}
{\bf -8.41 \ln{q^2\over m^2_t}} )\nonumber\\
&&
-(7.10\ln^2{q^2\over M^2_W}
+2.45\ln^2{q^2\over M^2_Z})
\}+........]\ ,
\label{sigbt}\eqa

\noindent
the cross section for
production of the five "light" ($u,d,s,c,b$) quarks $\sigma_5$ :

\bqa
\sigma_{5}&=&\sigma^{B}_{5}[1+{\alpha\over4\pi}\{(9.88N-42.66
)\ln{q^2\over\mu^2}\nonumber\\
&&+(46.58 \ln{q^2\over M^2_W}+7.25\ln{q^2\over M^2_Z}
{\bf -1.21 \ln{q^2\over m^2_t}})\nonumber\\
&&
-(6.30\ln^2{q^2\over M^2_W}
+2.03\ln^2{q^2\over M^2_Z})
\} +........]\ ,
\label{sig5t}
\eqa
\noindent
the forward-backward $b$-asymmetry

\bqa
A_{FB,b}&=&A^{B}_{FB,b}+{\alpha\over4\pi}\{(0.56N-6.13
)\ln{q^2\over\mu^2}\nonumber\\
&&+(17.23 \ln{q^2\over M^2_W}+0.96\ln{q^2\over M^2_Z}
{\bf -0.36 \ln{q^2\over m^2_t}})\nonumber\\
&&
-(0.31\ln^2{q^2\over M^2_W}
+0.08\ln^2{q^2\over M^2_Z})
\}+........ \ .
\label{AFBbt}\eqa
\noindent
For initial polarized leptons, we can consider
the longitudinal polarization $b$ asymmetry:

\bqa
A_{LR,b}&=&A^{B}_{LR,b}+{\alpha\over4\pi}\{(1.88N-20.46
)\ln{q^2\over\mu^2}\nonumber\\
&&+(27.91 \ln{q^2\over M^2_W}+1.92\ln{q^2\over M^2_Z}
{\bf -2.39 \ln{q^2\over m^2_t}})\nonumber\\
&&
-(2.35\ln^2{q^2\over M^2_W}+0.52\ln^2{q^2\over M^2_Z})\}+........ \ ,
\label{ALRbt}\eqa
\noindent
and the longitudinal polarization asymmetry for five light quark
production $A_{LR,5}$:

\bqa
A_{LR,5}&=&A^{B}_{LR,5}+{\alpha\over4\pi}\{(2.11N-22.95
)\ln{q^2\over\mu^2}\nonumber\\
&&+(24.07 \ln{q^2\over M^2_W}+1.63\ln{q^2\over M^2_Z}
{\bf -0.53 \ln{q^2\over m^2_t}})\nonumber\\
&&-(3.12\ln^2{q^2\over M^2_W}
 +0.55\ln^2{q^2\over M^2_Z})\}+........
\label{ALR5t}\eqa

In all eqs.(\ref{sigbt}-\ref{ALR5t}) the index $B$ refers to the
"Born" term computed with the "$Z$-peak subtracted" inputs, and
the dots on the r.h.s. correspond to
residual "non leading" asymptotic terms, that are either constant or
vanishing with $q^2$, whose role will be discussed later on. We have
written in "boldface" the genuine terms arising from the
contributions
$\simeq \alpha{m^2_t\over M^2_W}\ln{q^2\over m^2_t}$ discussed above.\par

For practical purposes it may be convenient to consider, rather than
$\sigma_b$, the ratio $R_b=\sigma_b/\sigma_5$, that generalizes the
analogous quantity defined at the $Z$ peak, For this observable we
would find the following expression:
\bqa
R_{b}&=&R^{B}_{b}[1+{\alpha\over4\pi}\{(N-11.16
)\ln{q^2\over\mu^2}\nonumber\\
&&+(30.17 \ln{q^2\over M^2_W}+4.73 \ln{q^2\over M^2_Z}
{\bf -7.20 \ln{q^2\over m^2_t}})
\nonumber\\
&& -(0.80\ln{q^2\over M^2_W}+0.42\ln{q^2\over M^2_Z})]
\}+........]
\label{Rbt}\eqa

Eqs.(\ref{sigbt}-\ref{Rbt}) are the main result of this paper. They
show in detail what is the extra effect of Sudakov-type coming from the
proper consideration of the actual top mass in the high energy regime (terms in 
boldface).
To give them a more quantitative meaning, we have computed the relative
top effect in the energy range between $1~TeV$ and $10~TeV$, where we
know from a previous analysis performed in ref.\cite{sud} that the
asymptotic logarithmic expansion (without the top effect) is well
reproducing the main features of an exact one-loop calculation. For
sake of completeness, we have also compared the relative top effects
with the \underline{overall} logarithmic effect (that also includes the
RG contributions).\par
The results of our investigation are shown in Table 1. 
>From inspection
of that Table, the following conclusions may be drawn:\par
a) The Sudakov top effect is sizeable in $R_b$ (this is practically due
to the $b$ cross section $\sigma_b$ in the numerator). At $1~TeV$, the
size of the relative shift is beyond 
the ``reference'' one percent limit; at 500 GeV,
the expected energy of the next LC, one
finds a relative effect of one percent. This
would be clearly visible at the aimed luminosity of the machine
\cite{tesla}.\par
b) in all four remaining observables, the Sudakov top effect is largely
diluted and, practically, hardly visible at an accuracy of few
permille. A possible exception to this statement might be offered by
the longitudinal polarization asymmetry $A_{LR,b}$ provided that
the available luminosity is of the few hundred $fb^{-1}$ size; this
might be kept in mind if longitudinal polarization became available.\par
c) The top contribution is systematically negative. Its inclusion
modifies the overall logarithmic expansion, but does not alter
dramatically its essential features. In other words, the discussion of
ref.\cite{sud} related to the order of magnitude of the various
logarithmic effects and of the total effect remains completely
valid.\par
One sees therefore a final picture that resembles very much the
corresponding situation already met at the $Z$ peak, where neglecting the
top contribution to $R_b$ from the "non oblique" diagram would have
been a theoretical disaster. This is, we believe, the main lesson that
may be learned from this paper.\par
To conclude this work, we want now to present, in the same spirit of
ref.\cite{sud}, an ``effective'' parametrization that describes the energy
dependence of the unpolarized observables $R_b$, $A_{FB,b}$ and
$\sigma_5$ in an energy region below $1~TeV$, where a priori the
asymptotic expansions might not be suitable. The conclusions of
ref.\cite{sud} were that, somehow surprizingly, the same
expressions that were derived for the asymptotic regime were able, with
the simple addition of a constant term, to reproduce the exact results
of the program TOPAZ0\cite{topaz0} within a few permille at most. In
the specific case of $\sigma_b$, $A_{FB,b}$, $\sigma_5$, we have thus
repeated the same procedure of comparison. This time, for
self-consistency reasons, we have limited our analysis to a range from
about $500~GeV$ to $1~TeV$, where the ratio ${q^2\over m^2_t}$ is (at
least) larger then ten.\par

A comparison of the logarithmic terms in the asymptotic 
expansion with the output of 
TOPAZ0 (of course, without inital state and 
final state QED corrections) gives 
the dashed curves shown in Fig.3. 
They are a rather poor approximation in this energy range, although 
they reproduce well the qualitative behaviour 
of the observables as functions of the energy. Motivated by 
the analysis in~\cite{sud}, we improve
the expansion by adding suitable constants $c_5$, $c_b$ 
and $c_{FB, b}$ according to the formulas:
\be
\sigma_{5, b} = \sigma^B_{5, b} ( 1 + \frac{\alpha}{\pi}(c_{5, b} 
+ \mbox{logarithms}))
\ee
\be
A_{FB, b} = A_{FB, b}^B + \frac{\alpha}{\pi}(c_{FB, b} 
+ \mbox{logarithms})
\ee
where, "logarithms" stands for the logarithms arising 
from the asymptotic
expansion and $\sigma_5^B$, $\sigma_b^B$, $A_{FB, b}^B$
are the Born level expressions of the observables.
With the following values
\be
c_5 = -28.9, \quad c_b = -26.7, \quad c_{FB, b} = -2.2 ,
\ee
we obtain the full lines in Fig.3. As in~\cite{sud}, the agreement is
very good and the percentual deviation is below the expected experimental
accuracy for all the three observables. In more details, $\sigma_5$ and $A_{FB, 
b}$ are 
precise at the 5 permille level; $\sigma_b$ displays a larger 1 \% deviation 
which 
can be explained in terms of the dominance of the top mass scale and of the 
large constant terms $\sim \ln m_t^2/M_W^2$.

As a final comment on our investigation, we would like to 
stress the following fact. At the considered one loop level, 
the linear logarithmic contribution of the top quark
can be considered as a ``subleading'' asymptotic effect, 
compared with the quadratic Sudakov logarithm.
As we have shown, the origin of this top term is of Yukawa 
type and thus proportional to 
$\alpha\frac{m_t^2}{M_W^2}\ln \frac{q^2}{m_t^2}$.

In a very recent and interesting paper~\cite{Kuhn}, the ``leading'' 
squared logarithms
of Sudakov type have been computed beyond the one loop 
approximation. In that analysis,
all the ``subleading'' single logarithmic contributions 
have been systematically
neglected. In particular, the top contribution that we have 
computed is not considered
in that approach. At the TeV energies on which we have concentrated 
our analysis, this contribution,
although of ``subleading'' kind, is nonetheless quite relevant and 
cannot be neglected.

The same qualitative remark should apply in our opinion to
 \underline{all} the ``subleading''
logarithmic terms. At the one loop level, the complete 
calculation of~\cite{sud} shows that
the numerical weights of the linear Sudakov logarithms, 
in the TeV region, is of the same
size (with opposite sign) as that of the ``leading'' 
quadratic logarithms. This makes
us feel that, to obtain a fully satisfactory prediction, the 
calculation of \underline{all}
``subleading'' logarithms beyond the one loop level should be 
included. 

\newpage

\begin{center}
{\bf Table 1:
Logarithmic contributions to various observables}\\
(1) without top effects ($b\equiv d\equiv s$); (2) with top effects;
(3) top effects alone.\\
\vspace*{0.5cm}
\begin{tabular}{|c|c|c|c|c|c|c|}
\hline
\multicolumn{1}{|c|}{$\sqrt{q^2}~(TeV)$}&
\multicolumn{1}{|c|}{$\delta\sigma_b/\sigma_b$} &
\multicolumn{1}{|c|}{$\delta R_b/R_b$} &
\multicolumn{1}{|c|}{$\delta\sigma_5/\sigma_5$} &
\multicolumn{1}{|c|}{$\delta A_{FB,b}$}&
\multicolumn{1}{|c|}{$\delta A_{LR,b}$}&
\multicolumn{1}{|c|}{$\delta A_{LR,5}$}
\\[0.1cm] \hline
$0.3 ~~~(1)$&
0.0723 &  0.0391 &  0.0332 &  0.0213 &  0.0145  & 0.0018\\
$0.3 ~~~(2)$&
 0.0667  & 0.0344 &  0.0324 &  0.0211 &  0.0130  & 0.0015\\
$0.3 ~~~(3)$&
-0.0056&  -0.0048 & -0.0008 & -0.0002 & -0.0016 & -0.0004\\
[0.1cm] \hline
$0.5 ~~~(1)$&
0.0778 &  0.0513 &  0.0265 &  0.0284 &  0.0128 & -0.0069\\
$0.5 ~~~(2)$&
0.0669 &  0.0420 &  0.0250 &  0.0280 &  0.0097 & -0.0075\\
$0.5~~ ~(3)$&
-0.0109 & -0.0093 & -0.0016 & -0.0005 & -0.0031 & -0.0007\\
[0.1cm] \hline
$1 ~~~(1)$&
 0.0656  & 0.0653  & 0.0003 &  0.0373 &  0.0045 & -0.0262\\
$1 ~~~(2)$&
       0.0475 &  0.0498 & -0.0023 &  0.0366 & -0.0007 & -0.0273\\
$1~~ ~(3)$&
-0.0181 & -0.0155 & -0.0026 & -0.0008 & -0.0052 & -0.0011\\ 
[0.1cm] \hline 
$2~~ ~(1)$&
       0.0307 &  0.0764 & -0.0457 &  0.0453 & -0.0106 & -0.0542\\
$2 ~~~(2)$&
       0.0054 &  0.0547 & -0.0493 &  0.0442 & -0.0178 & -0.0558\\
$2 ~~~(3)$&
 -0.0253 & -0.0217 & -0.0036 & -0.0011&  -0.0072 & -0.0016\\
[0.1cm] \hline
$5 ~~~(1)$&
      -0.0502 &  0.0866 & -0.1368 &  0.0544 & -0.0411 & -0.1046\\
$5 ~~~(2)$&
       -0.0851 &  0.0568 & -0.1419 &  0.0529 & -0.0510 & -0.1068\\
$5 ~~~(3)$&
 -0.0348 & -0.0298 & -0.0050 & -0.0015 & -0.0099 & -0.0022\\
[0.1cm] \hline
$10 ~~~(1)$&
      -0.1377 &  0.0910 & -0.2288 &  0.0602 & -0.0720 & -0.1529\\
$10 ~~~(2)$&
      -0.1798 &  0.0550 & -0.2348 &  0.0584 & -0.0839 & -0.1555\\
$10 ~~ ~(3)$&
  -0.0420 & -0.0360 & -0.0061 & -0.0018&  -0.0119 & -0.0027\\
[0.1cm] \hline
\end{tabular}\\
\end{center}

\begin{figure}[p]
\[
\hspace{-2.5cm}
\epsfig{file=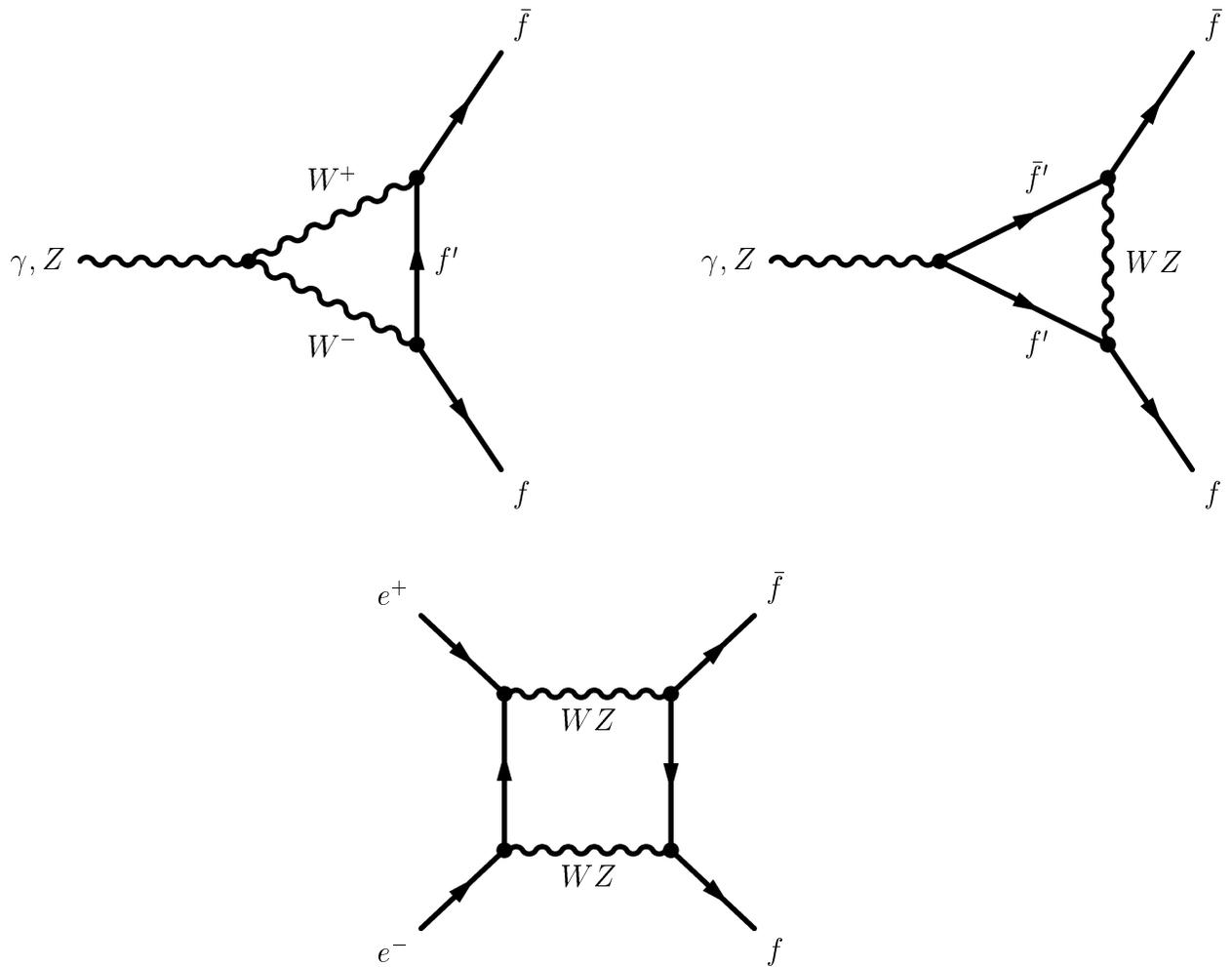,height=30cm}
\]
\vspace*{-11cm}
\caption[1]{Triangle and box diagrams for final light quark pairs.}
\label{diag1}
\end{figure}

\begin{figure}[p]
\[
\hspace{-3cm}
\epsfig{file=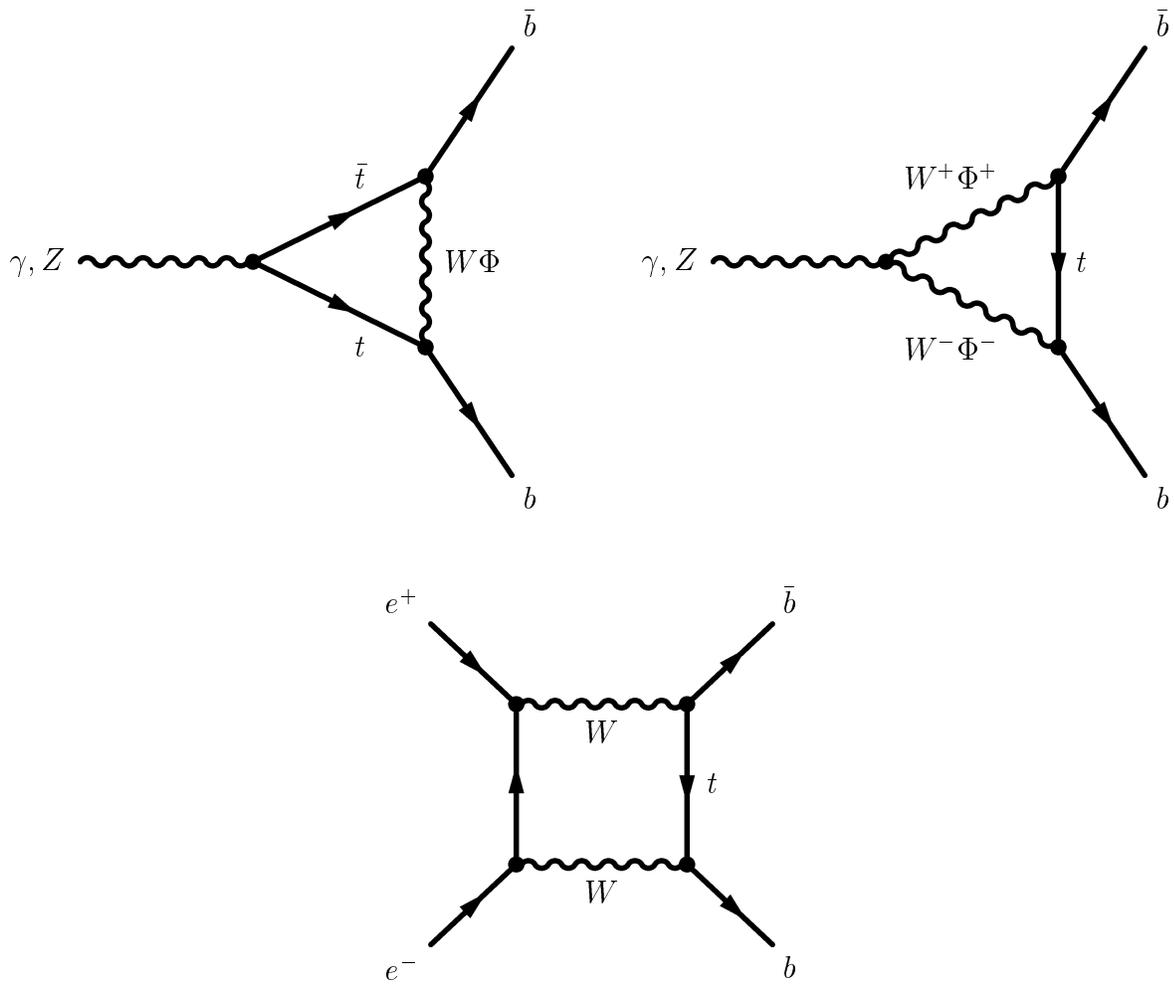,height=30cm}
\]
\vspace*{-11cm}
\caption[2]{Triangle and box diagrams for final $b\bar b$
quarks. $\Phi^{\pm}$ are the charged would-be Goldstone
bosons.}
\label{diag2}
\end{figure}

\begin{figure}[p]
\[
\epsfig{file=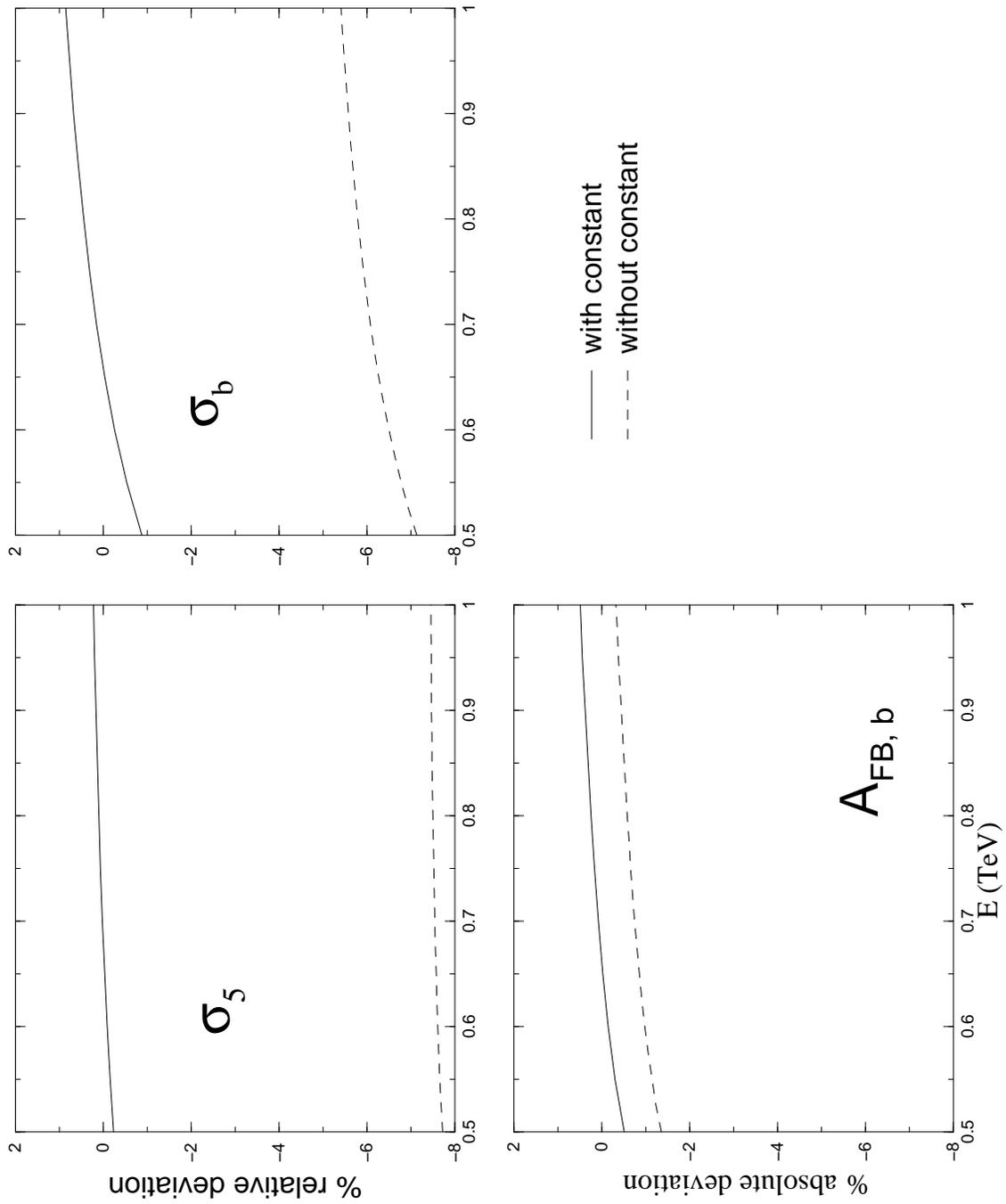,height=18cm}
\]
\caption[2]{Comparison between TOPAZ0 and the logarithmic expansion. For the two cross sections 
$\sigma_5$ and $\sigma_b$ we show the relative percentual deviation with and without
constant terms. In the case of the asymmetry $A_{FB, b}$ the shown deviation is absolute and not relative.}
\label{topaz0}
\end{figure}

\end{document}